\documentclass[preprint,review,12pt]{elsarticle}
\biboptions{sort&compress}

\usepackage[T2A,T1]{fontenc}
\usepackage{amsmath,amssymb}
\usepackage[absolute]{textpos}

\begin{document}
\begin{textblock*}{\paperwidth}(1.5in,0.5in)
\footnotesize\noindent\copyright\ 2023. This manuscript version is made available under the CC-BY-NC-ND 4.0 license

\noindent https://creativecommons.org/licenses/by-nc-nd/4.0/
\end{textblock*}
\begin{frontmatter}

\title{Perfect nonradiating electromagnetic source and its self-action}

\author{Denis Nikolaevich Sob'yanin \fontencoding{T2A}\selectfont(Денис Николаевич Собьянин)\fontencoding{T1}\selectfont}
\ead{sobyanin@lpi.ru}
\affiliation{organization={I. E. Tamm Division of Theoretical Physics, P. N. Lebedev Physical Institute of the Russian Academy of Sciences},
            addressline={Leninskii Prospekt 53},
            city={Moscow},
            postcode={119991},
            country={Russia}}

\begin{abstract}
Nonradiating sources and anapoles are curious objects from the physics of invisibility that illuminate subtle concepts of fundamental electrodynamics and have promising applications in nanophotonics. The present work shows that a perfectly conducting sphere with a hidden magnetic field is a simple nonradiating electromagnetic source with mechanical excitation and complete internal confinement of electromagnetic energy. It does not require external electromagnetic excitation and is excited by rotation, which induces internal charges and currents with a self-compensating external radiation. The constructed source acts on itself through Lorentz forces emerging from the interaction of charges and currents with electromagnetic fields inside the sphere and, when having a freedom of rotation about the fixed center, exhibits regular precession like a gyroscope. This self-action reveals an internal electromagnetic activity of the perfect nonradiating source, externally inactive and invisible. Neutron stars, these nanospheres of space, can be such natural nonradiating sources when their magnetic fields are buried.
\end{abstract}

\end{frontmatter}

\section{Introduction}

The history of nonradiating sources of electromagnetic radiation, famous for their self-contradictory name and unusual properties, began more than 100 years ago \cite{Gbur2003}. It goes back to the first attempts to explain the stability of atoms from the classical viewpoint before the invention of quantum mechanics \cite{Ehrenfest1910}. The common belief having motivated the development of new quantum ideas was that accelerated charges must radiate. However, it has appeared that this situation does not always take place, and various types of nonradiating charge distributions in accelerated motion were found with thinking of probable applications in particle physics \cite{Schott1933,BohmWeinstein1948,Goedecke1964,AbneyGbur2023}. In the same field the term ``anapole'' was originally introduced for a magnetic analog to a spherical capacitor representing a bent solenoid without external magnetic field \cite{Zeldovich1958}. Now this term usually refers to a different system consisting of two dynamic electric and toroidal dipoles excited in such a way that their far radiation fields destructively interfere \cite{AfanasievStepanovsky1995}. Due to their nonradiating properties such systems resemble atoms and are often named meta-atoms \cite{ZanganehEtal2021}, the more so that their excitation is observed within various engineered metamaterials \cite{KaelbererEtal2010,FedotovEtal2013,BhattacharyaEtal2022}, a standard platform for studying nonstandard effects of light-matter interaction \cite{Veselago1968,XiaoEtal2020,DinhEtal2021}. An important property of confining and enhancing electromagnetic fields gives a promising base for absorption engineering and compact energy transfer at the nanoscale \cite{HuangEtal2021,HeEtal2022,QuEtal2022}.

The excitation of nonradiating systems is usually achieved through an external electromagnetic illumination that induces the needed anapole-like response in a given system \cite{MiroshnichenkoEtal2015,ParkerEtal2020,DeyEtal2022}. In this case we have two conceptually different electromagnetic fields: the inducing field that causes the system to be an electromagnetic source and the induced field that is generated by the thus excited source. However, the two fields do not differ physically and are in fact related through the system. A question naturally arises whether it is possible to construct, at least on a conceptual level, such a nonradiating source that has external electromagnetic fields belonging exclusively to it itself and is not excited by external radiation or other electromagnetic means. In this paper we will give a positive answer to this question by presenting a conceptually simple nonradiating source that is excited not electromagnetically but mechanically. It appears to be self-acting, which reflects the generation of internal electromagnetic fields inside the source and indicates the intimate intertwining of its optomechanical properties in spite of the absence of outgoing radiation.

\section{Results and Discussion}

\subsection{Nonradiation via the External Field Compensation}

One of the simplest sources of radiation is a magnetic dipole. A changing magnetic moment $\mathbf{m}$ radiates with total intensity $I=(2/3)\ddot{\mathbf{m}}^2/c^3$, where overdot denotes time derivative. The dipole may be constant in magnitude and then change its direction only, i.e., somehow rotate, $\dot{\mathbf{m}}=\mathbf{\Omega}\times\mathbf{m}$, where $\mathbf{\Omega}$ is an angular velocity. A possible realization of such a source is a rotating uniformly magnetized perfectly conducting sphere. The magnetic moment $\mathbf{m}=\mathbf{B} R^3/2$ corresponds to the sphere with radius $R$ and internal uniform magnetic field $\mathbf{B}$. In astrophysics, the rotating magnetized sphere is the basic model of the pulsar and by equating the electromagnetic and mechanical energy loss allows one to estimate its magnetic field \cite{Lipunov1992}.

The aforementioned sphere is the source of charges and currents that are responsible for the radiation. These charges and currents are generated mechanically through rotation of the sphere, and the generation is similar to the effect of unipolar induction in the Faraday disk, when rotation generates electromotive force. A more convenient way to understand how the sphere radiates is that outside the sphere, i.e., in the area $r>R$, where $r$ is the radial coordinate, we have an electromagnetic field that satisfies the source-free Maxwell equations, the Sommerfeld radiation condition at infinity, and the boundary conditions at the surface of the sphere, viz., the continuity of tangential component of electric field $\mathbf{E}$ and normal component of magnetic field $\mathbf{B}$,
\begin{equation}
\label{ntimesE}
\mathbf{n}\times\mathbf{E}_\text{ex}=\mathbf{n}\times\mathbf{E}_\text{in}\text{ at }r=R,
\end{equation}
\begin{equation}
\label{ndotB}
\mathbf{n}\cdot\mathbf{B}_\text{ex}=\mathbf{n}\cdot\mathbf{B}_\text{in}\text{ at }r=R,
\end{equation}
where $\mathbf{E}_\text{ex}$, $\mathbf{B}_\text{ex}$ and $\mathbf{E}_\text{in}$, $\mathbf{B}_\text{in}$ are the external and internal electric and magnetic fields, respectively, $\mathbf{r}$ is the radius vector, and $\mathbf{n}=\mathbf{r}/r$ is the radial unit vector. The internal electric and magnetic fields are related by the condition of perfect conductivity
\begin{equation}
\label{EminusvB}
\mathbf{E}_\text{in}=-\frac1c\mathbf{v}\times\mathbf{B}_\text{in},
\end{equation}
where
\begin{equation}
\label{vOmegar}
\mathbf{v}=\mathbf{\Omega}\times\mathbf{r}
\end{equation}
is the velocity of the conducting medium at a given point $\mathbf{r}$ of the sphere rigidly rotating with angular velocity $\mathbf{\Omega}$. Rotation of the magnetized sphere, when we have nonzero $\mathbf{v}$ and $\mathbf{B}$, results in the appearance of changing electromagnetic fields just under the surface $r=R$ and, correspondingly, changing boundary conditions. Thus, the mechanism of radiation is that the tangential electric and normal magnetic field components at the surface $r=R$ existing and changing because of the rotation determine nonzero and changing external electromagnetic fields outside the sphere.

\begin{figure}
\centering
\includegraphics[width=8cm]{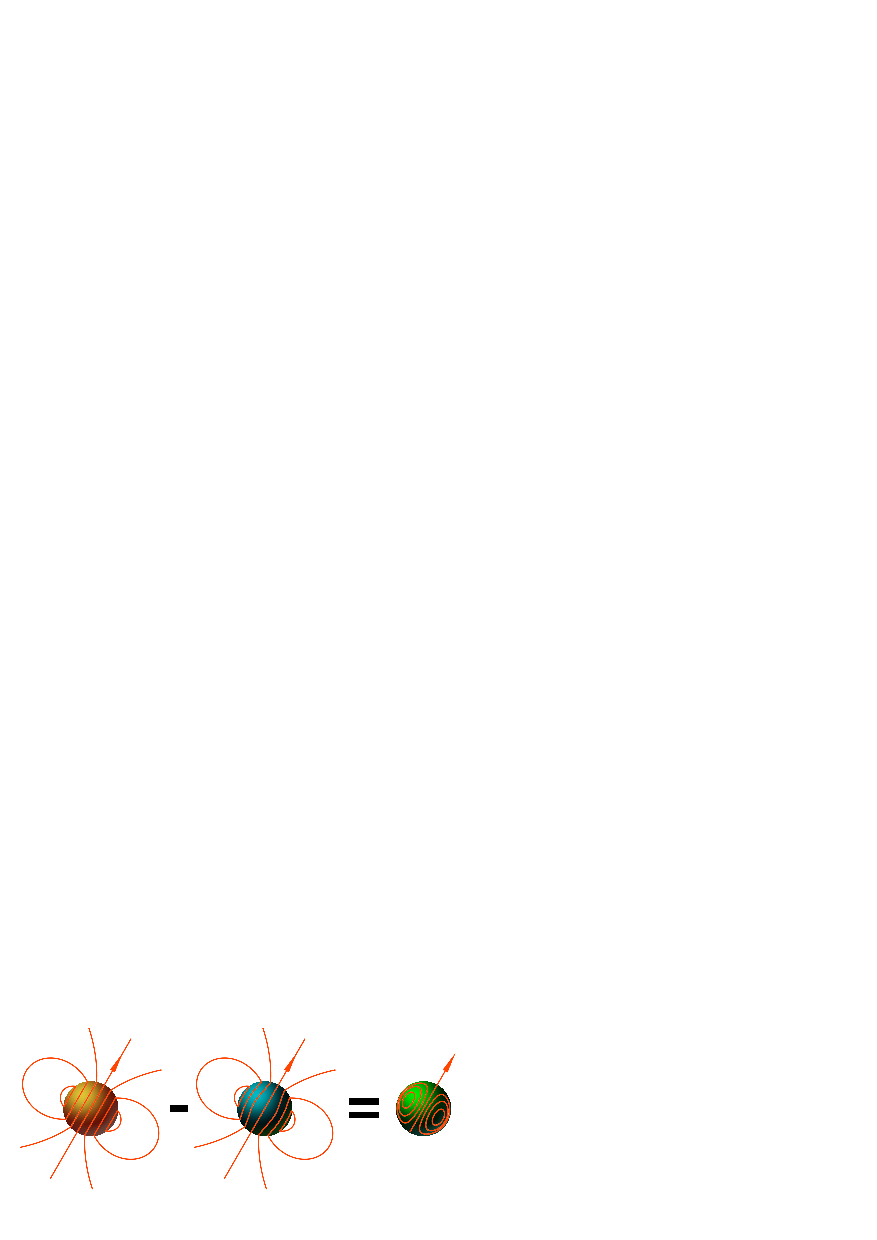}
\caption{Cancellation of the external electromagnetic fields.}
\label{fig1}
\end{figure}

Is it possible to construct a nonradiating electromagnetic source on the basis of a rotating magnetized sphere? The inverse electromagnetic problem is nonunique~\cite{BleisteinCohen1977}, which means that different sources may give the same radiation field. In this relation recall the aforementioned idea used for the anapole, a combination of the toroidal dipole and the usual electric dipole: the two dipoles generate the same electromagnetic field in the far zone, which through proper synchronization of the dipoles leads to the destructive interference in the anapole field. Returning to our case, we will construct an alternative electromagnetic source that generates the same electromagnetic field as the described rotating uniformly magnetized sphere and then combine the two sources to obtain nonradiation (Fig.~\ref{fig1}).

\subsection{Electromagnetic Field Distribution}

As follows from the above description of the field generation by the sphere, any source with the same boundary conditions will give the same external electromagnetic field, so it is sufficient to consider a conducting sphere with some internal magnetic field different from a constant $\mathbf{B}$ but providing the same boundary conditions at $r=R$ as the rotating uniformly magnetized sphere. The internal magnetic field can be represented through a vector potential $\mathbf{A}_\text{in}$, $\mathbf{B}_\text{in}=\nabla\times\mathbf{A}_\text{in}$, which for uniform magnetization is $\mathbf{A}=\mathbf{B}\times\mathbf{r}/2$, so that $\mathbf{B}=\nabla\times\mathbf{A}$. For an alternative source we may take a simple vector potential $\mathbf{A}_0=(r/R)\mathbf{A}$ that differs from $\mathbf{A}$ inside the sphere but coincides with $\mathbf{A}$ at the boundary. The internal magnetic field $\mathbf{B}_0=\nabla\times\mathbf{A}_0$ becomes
\begin{equation}
\label{B0}
\mathbf{B}_0=\frac{r}{2R}(3\mathbf{B}-\mathbf{n}\mathbf{n}\cdot\mathbf{B}),
\end{equation}
and since $\mathbf{n}\cdot\mathbf{B}_0=(r/R)\mathbf{n}\cdot\mathbf{B}$, boundary condition \eqref{ndotB} is the same for both sources, $\mathbf{n}\cdot\mathbf{B}_0=\mathbf{n}\cdot\mathbf{B}$ at $r=R$. Note that throughout the paper we use the dyadic notations, so that $\mathbf{a}\mathbf{b}=||a_i b_j||$ and $\mathbf{a}\mathbf{b}\cdot\mathbf{c}=\mathbf{a}(\mathbf{b}\cdot\mathbf{c})$, $\mathbf{a}\cdot\mathbf{b}\mathbf{c}=(\mathbf{a}\cdot\mathbf{b})\mathbf{c}$, $\mathbf{a}\cdot\mathbf{b}\mathbf{c}\cdot\mathbf{d}=(\mathbf{a}\cdot\mathbf{b})(\mathbf{c}\cdot\mathbf{d})$, and $\mathbf{a}\cdot\mathbf{b}\mathbf{c}\times\mathbf{d}=(\mathbf{a}\cdot\mathbf{b})(\mathbf{c}\times\mathbf{d})$ for vectors $\mathbf{a}$, $\mathbf{b}$, $\mathbf{c}$, and $\mathbf{d}$. The corresponding internal electric field is calculated using Eqs.~\eqref{EminusvB} and~\eqref{vOmegar},
\begin{equation}
\label{E}
\mathbf{E}=\frac{r}{c}(\mathbf{\Omega}\mathbf{n}\cdot\mathbf{B}-\mathbf{n}\mathbf{\Omega}\cdot\mathbf{B})
\end{equation}
for the uniformly magnetized sphere with $\mathbf{B}_\text{in}=\mathbf{B}$ and
\begin{equation}
\label{E0}
\mathbf{E}_0=\frac{r^2}{2cR}[2\mathbf{\Omega}\mathbf{n}\cdot\mathbf{B}+\mathbf{n}(\mathbf{\Omega}\cdot\mathbf{n}\mathbf{n}\cdot\mathbf{B}-3\mathbf{\Omega}\cdot\mathbf{B})]
\end{equation}
for the equivalent electromagnetic source with $\mathbf{B}_\text{in}=\mathbf{B}_0$. Since $\mathbf{n}\times\mathbf{E}=-(1/c)\mathbf{v}\mathbf{n}\cdot\mathbf{B}$ and $\mathbf{n}\times\mathbf{E}_0=-(r/cR)\mathbf{v}\mathbf{n}\cdot\mathbf{B}$, boundary condition \eqref{ntimesE} is also the same for both sources, $\mathbf{n}\times\mathbf{E}_0=\mathbf{n}\times\mathbf{E}$ at $r=R$.

\begin{figure}
\centering
\includegraphics[width=8cm]{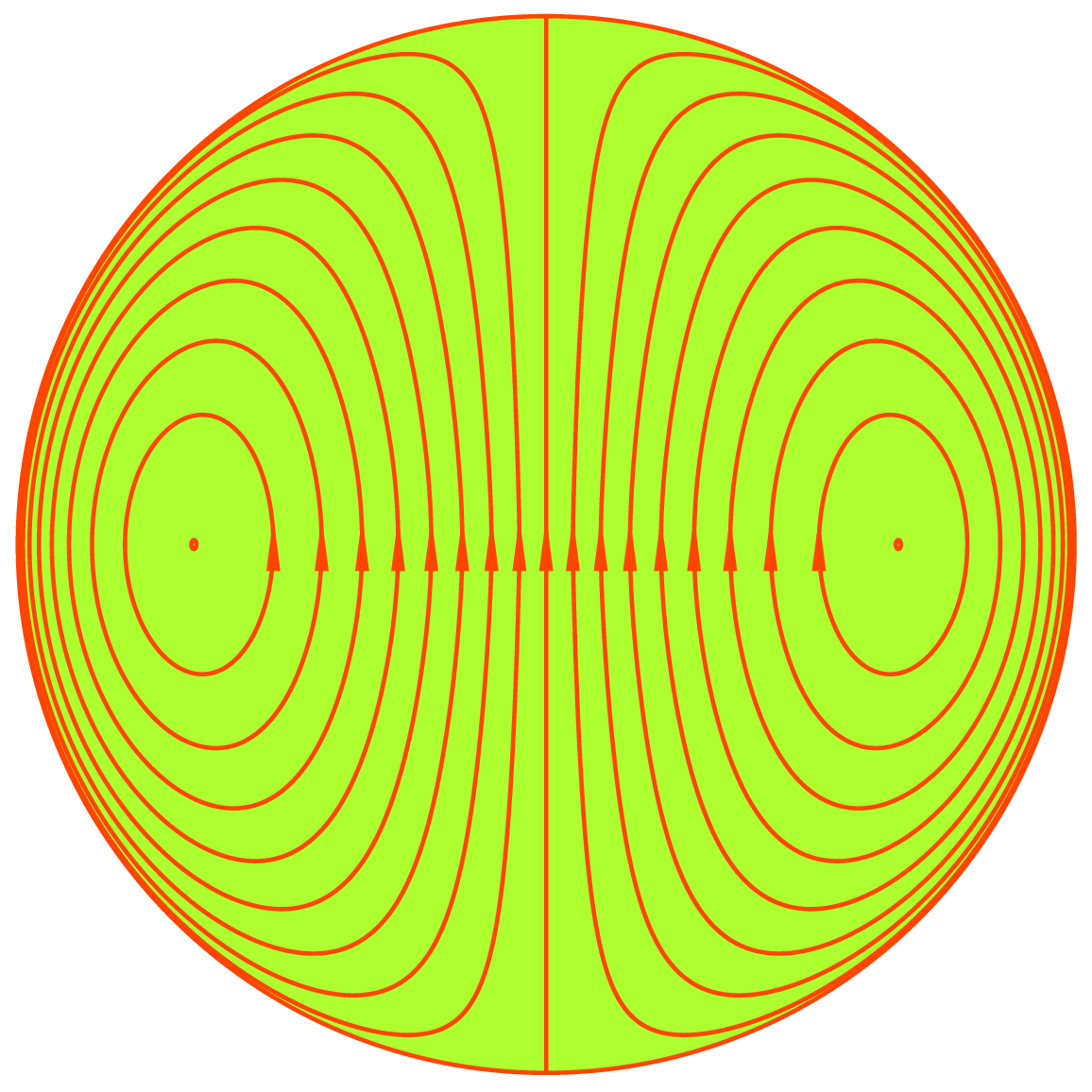}
\caption{Hidden magnetic field in the nonradiating sphere.}
\label{fig2}
\end{figure}

We see that two identically rotating perfectly conducting spheres of the same radius but with different internal magnetic fields $\mathbf{B}$ and $\mathbf{B}_0$ provide equivalent boundary conditions at their surfaces, so the electromagnetic field radiated by these spheres is the same at every point of the area $r>R$. To construct a nonradiating source, it remains to subtract the two electromagnetic fields in the whole space. Thus, an electromagnetic source representing a perfectly conducting rigidly rotating sphere with internal magnetic field
\begin{equation}
\label{Bin}
\mathbf{B}_\text{in}=\mathbf{B}-\mathbf{B}_0,
\end{equation}
where $\mathbf{B}_0$ is determined by Eq.~\eqref{B0}, does not radiate. This magnetic field is presented in Fig.~\ref{fig2} and, together with $\mathbf{B}$ and~$\mathbf{B}_0$, satisfies \cite{Sobyanin2016}
\begin{equation}
\label{dBindt}
\dot{\mathbf{B}}_\text{in}=\frac{\partial\mathbf{B}_\text{in}}{\partial t}+\mathbf{v}\cdot\nabla\mathbf{B}_\text{in}=\mathbf{\Omega}\times\mathbf{B}_\text{in},
\end{equation}
which means that the magnetic field distribution is frozen into the medium and rotates with angular velocity $\mathbf{\Omega}$ as a whole synchronously with the sphere. The rotationally induced internal electric field is
\begin{equation}
\label{Ein}
\mathbf{E}_\text{in}=\mathbf{E}-\mathbf{E}_0,
\end{equation}
where $\mathbf{E}$ and $\mathbf{E}_0$ are determined by Eqs.~\eqref{E} and~\eqref{E0}.

\subsection{Nonradiating Time-Dependent Charge and Current Distributions}

The constructed nonradiating electromagnetic source has nonzero volume charge and current densities in the area $r<R$,
\begin{equation}
\label{rho}
\rho=\frac1{4\pi c}\biggl[-2\mathbf{\Omega}\cdot\mathbf{B}+\frac{r}{R}(5\mathbf{\Omega}\cdot\mathbf{B}-3\mathbf{\Omega}\cdot\mathbf{n}\mathbf{n}\cdot\mathbf{B})\biggr]
\end{equation}
and
\begin{equation}
\label{j}
\begin{split}
\mathbf{j}=\frac1{4\pi c}\biggl[&-\frac{2c^2}{R}\mathbf{n}\times\mathbf{B}+\mathbf{\Omega}\mathbf{v}\cdot\mathbf{B}
\\
&-\frac{r}{R}\mathbf{v}\cdot\mathbf{B}\biggl(\mathbf{\Omega}+\frac12\mathbf{n}\mathbf{n}\cdot\mathbf{\Omega}\biggr)+\mathbf{w}\times\mathbf{B}_\text{in}\biggr],
\end{split}
\end{equation}
where
\begin{equation}
\label{w}
\mathbf{w}=\dot{\mathbf{\Omega}}\times\mathbf{r}
\end{equation}
is the rotational acceleration. Densities \eqref{rho} and \eqref{j} satisfy the standard charge conservation law $\partial\rho/\partial t+\nabla\cdot\mathbf{j}=0$. In addition to these densities, at $r=R$ there also appear the surface charge and current densities
\begin{equation}
\label{rhoSigma}
\rho_\text{surf}=\frac{R}{8\pi c}(\mathbf{\Omega}\cdot\mathbf{n}\mathbf{n}\cdot\mathbf{B}-\mathbf{\Omega}\cdot\mathbf{B})
\end{equation}
and
\begin{equation}
\label{jSigma}
\mathbf{j}_\text{surf}=\frac{c}{8\pi}\mathbf{n}\times\mathbf{B},
\end{equation}
which satisfy the charge conservation law $\partial\rho_\text{surf}/\partial t+\nabla_\text{surf} \cdot\mathbf{j}_\text{surf}=\mathbf{n}\cdot\mathbf{j}|_{r=R}$, with $\nabla_\text{surf} \cdot\mathbf{j}_\text{surf}$ being the two-dimensional divergence at the surface of the sphere, meaning that the increase in charge in a given surface element equals the volume charge flow from inside the sphere minus the surface charge flow through the boundary of the element.

Thus, nonzero volume and surface charges and currents determined by Eqs.~\eqref{rho}--\eqref{jSigma} and bounded in the area $r\leqslant R$ generate such electromagnetic fields that fully compensate each other in the area $r>R$ and lead to strict nonradiation. These charges and currents are induced purely mechanically, through rotation of the sphere, and do not require any external electromagnetic sources, such as wires, capacitors, and coils. It is not required that the charges and currents be harmonic with time dependence $\propto e^{-i\Omega t}$ for a constant $\Omega$, which means that rotation of the sphere may be arbitrary and both the absolute value and the direction of angular velocity $\mathbf{\Omega}$ may change in time.

\subsection{Self-Torque and Precession}

An important question arises as to whether the constructed source differs in any aspect (beyond the very fact of nonzero internal charges and currents) from, say, a trivial nonmagnetized conducting sphere of the same radius $R$ that has no charges and currents at all. Both spheres have zero electromagnetic fields outside and cannot be distinguished by external radiation or energy dissipation. However, the constructed source appears to demonstrate an interesting effect of self-action: rotation of the sphere does not generate external electromagnetic fields but generates nonzero internal electromagnetic fields, and the interaction of internal charge and current densities with these fields results in nonzero Lorentz forces that act on our source. The torque
\begin{equation}
\label{M}
\mathbf{M}=\mathbf{M}_\text{vol}+\mathbf{M}_\text{surf}
\end{equation}
of these Lorentz forces is the sum of volume and surface parts,
\begin{equation}
\label{Mv}
\mathbf{M}_\text{vol}=\int\mathbf{r}\times\biggl(\rho\mathbf{E}_\text{in}+\frac1c \mathbf{j}\times\mathbf{B}_\text{in}\biggr)\,dV
\end{equation}
and
\begin{equation}
\label{Msigma}
\mathbf{M}_\text{surf}=\int\mathbf{r}\times\biggl(\rho_\text{surf}\mathbf{E}_\text{in}+\frac1c \mathbf{j}_\text{surf}\times\mathbf{B}_\text{in}\biggr)\,dS,
\end{equation}
with integration in Eq.~\eqref{Mv} over the volume and in Eq.~\eqref{Msigma} over the surface or the sphere. Substituting Eqs.~\eqref{Bin}, \eqref{Ein}, \eqref{rhoSigma}, and \eqref{jSigma} into Eq.~\eqref{Msigma}, we have that the surface torque vanishes,
\begin{equation}
\label{Msigma0}
\mathbf{M}_\text{surf}=0.
\end{equation}

To grasp the existence of self-action, first consider the rotation of our sphere with constant angular velocity $\mathbf{\Omega}$, i.e., the uniform rotation about some fixed axis. In this case acceleration \eqref{w} is zero and the resulting torque becomes
\begin{equation}
\label{MOmega}
\mathbf{M}_\Omega=-\frac{11}{1260}\frac{R^5}{c^2}\mathbf{\Omega}\cdot\mathbf{B}\mathbf{\Omega}\times\mathbf{B}.
\end{equation}
We see that the torque is orthogonal to the rotational axis, trying to rotate the sphere about this orthogonal direction, and only the fixed axis does not allow this extra rotation but it is subjected to the corresponding loading. By contrast, the nonmagnetized sphere of uniform mass density will freely rotate about any fixed axis passing through its center not loading the axis.

Now consider what will happen if the fixed axis is removed and does not prevent the torque from changing the initial motion of the sphere. For a generally changing angular velocity $\mathbf{\Omega}$ the part $\propto\mathbf{w}\times\mathbf{B}_\text{in}$ begins to play the role in Eq.~\eqref{j} and correspondingly in Eq.~\eqref{Mv}, and the total torque
\begin{equation}
\label{M1}
\mathbf{M}=\mathbf{M}_\Omega+\mathbf{M}_{\dot{\Omega}}
\end{equation}
will contain not only part \eqref{MOmega} depending on $\mathbf{\Omega}$ but also an extra part depending on~$\dot{\mathbf{\Omega}}$,
\begin{equation}
\label{MOmegaDot}
\mathbf{M}_{\dot{\Omega}}=-\frac{1}{1260}\frac{R^5}{c^2}(5B^2\dot{\mathbf{\Omega}}+11\dot{\mathbf{\Omega}}\cdot\mathbf{B}\mathbf{B}).
\end{equation}
When the rotational axis of the sphere is not fixed, the torque may change the vector of angular velocity, but this change in turn itself changes the initial torque through Eq.~\eqref{MOmegaDot}. This means that the sphere rotates in a self-consistent way and generates during its rotation such a torque that leads to the initial rotation. The equation of motion for the sphere with the moment of inertia $J$ is
\begin{equation}
\label{EoM}
J\dot{\mathbf{\Omega}}=\mathbf{M},
\end{equation}
where $\mathbf{M}$ is determined from Eq.~\eqref{M1}.

Scalarly multiplying Eq.~\eqref{EoM} by $\mathbf{B}$ gives $[J+(4/315)R^5B^2/c^2]\dot{\mathbf{\Omega}}\cdot\mathbf{B}=0$. Therefore,
\begin{equation}
\label{OmegaDotDotB}
\dot{\mathbf{\Omega}}\cdot\mathbf{B}=0
\end{equation}
and Eq.~\eqref{EoM} transforms to
\begin{equation}
\label{EoM1}
J_\text{eff}\dot{\mathbf{\Omega}}=\mathbf{M}_\Omega,
\end{equation}
where we have introduced an effective moment of inertia
\begin{equation}
J_\text{eff}=J+\frac1{252}\frac{R^5}{c^2}B^2.
\end{equation}
Scalarly multiplying Eq.~\eqref{EoM1} by $\mathbf{\Omega}$ gives $\dot{\mathbf{\Omega}}\cdot\mathbf{\Omega}=0$ and hence
\begin{equation}
\label{OmegaConst}
\Omega=const,
\end{equation}
so the angular velocity $\mathbf{\Omega}$ in fact changes only in direction and not in magnitude. Since analogously to Eq.~\eqref{dBindt}
\begin{equation}
\label{dBdt}
\dot{\mathbf{B}}=\mathbf{\Omega}\times\mathbf{B},
\end{equation}
Eq.~\eqref{OmegaDotDotB} implies $(\mathbf{\Omega}\cdot\mathbf{B})\dot{}=\dot{\mathbf{\Omega}}\cdot\mathbf{B}+\mathbf{\Omega}\cdot\dot{\mathbf{B}}=0$ and
\begin{equation}
\label{OmegaDotB0}
\mathbf{\Omega}\cdot\mathbf{B}=const,
\end{equation}
which together with Eq.~\eqref{OmegaConst} indicates that the angle between $\mathbf{\Omega}$ and $\mathbf{B}$ does not change.

It follows from Eqs.~\eqref{MOmega}, \eqref{EoM1}, and \eqref{OmegaDotB0} that $[J_\text{eff}\mathbf{\Omega}+(11/1260)R^5\mathbf{\Omega}\cdot\mathbf{B}\mathbf{B}/c^2]\dot{}=0$, so we may define a vector
\begin{equation}
\label{OmegaPrec}
\mathbf{\Omega}_\text{prec}=\mathbf{\Omega}-\mathbf{\Omega}_\text{prop}=const
\end{equation}
that is constant both in magnitude and in direction while a vector
\begin{equation}
\label{OmegaProp}
\mathbf{\Omega}_\text{prop}=-\frac{11}{1260}\frac{R^5}{c^2 J_\text{eff}}\mathbf{\Omega}\cdot\mathbf{B}\mathbf{B}
\end{equation}
is constant only in magnitude, as follows from Eq.~\eqref{OmegaDotB0},
\begin{equation}
\label{OmegaPropConst}
\Omega_\text{prop}=const.
\end{equation}
Vectorially multiplying Eq.~\eqref{OmegaPrec} by $\mathbf{\Omega}$ and comparing the result to Eq.~\eqref{EoM1}, we get
\begin{equation}
\label{dOmegadt}
\dot{\mathbf{\Omega}}=\mathbf{\Omega}_\text{prec}\times\mathbf{\Omega}
\end{equation}
and correspondingly
\begin{equation}
\label{dOmegaPropdt}
\dot{\mathbf{\Omega}}_\text{prop}=\mathbf{\Omega}_\text{prec}\times\mathbf{\Omega}_\text{prop}.
\end{equation}
Equations \eqref{dOmegadt} and \eqref{dOmegaPropdt} mean that both vectors $\mathbf{\Omega}$ and $\mathbf{\Omega}_\text{prop}$ retain their angles with respect to the constant direction of $\mathbf{\Omega}_\text{prec}$ and rotate uniformly about it, so $\mathbf{\Omega}_\text{prec}$ appears to be the angular velocity of precession, while $\mathbf{\Omega}_\text{prop}$, the difference between $\mathbf{\Omega}$ and~$\mathbf{\Omega}_\text{prec}$, corresponds to the angular velocity of proper rotation. Thus, the electromagnetic self-action of our nonradiating sphere reveals itself in regular precession, the motion characteristic of gyroscopes \cite{BorisovMamaev2019}: the sphere rotates with constant angular frequency $\Omega_\text{prop}$ about the axis fixed in the sphere and determined by the direction of $\mathbf{B}$ while the axis itself rotates with constant angular frequency $\Omega_\text{prec}$ about the axis fixed in space and determined by the direction of~$\mathbf{\Omega}_\text{prec}$. We see that though any outgoing radiation or externally applied fields, which could affect the mechanical dynamics, are absent, the dynamics of the sphere is changed, and the internal electromagnetic fields are the cause. This effect can be considered as an uncommon reflection of the general optomechanical coupling.

\subsection{Natural Realization}

Can the proposed nonradiating source exist in nature? Not speculating on how to artificially create something similar in the laboratory, note instead an interesting mechanism that might create something similar in space. We have already mentioned that pulsars, cosmic objects representing neutron stars with a radius of 10~km and very strong magnetic fields reaching $3\times10^{16}$~G \cite{Sobyanin2023} can be considered in the simplest case as rotating magnetized spheres. When these are born in supernova explosions, the accretion of the conducting medium onto the neutron star can bury the external magnetic field so that the magnetic field at the surface of the star vanishes \cite{GeppertEtal1999}. Thus, neutron stars with buried magnetic fields can in fact be natural nonradiating sources, though the structure of these fields is unknown and may differ from the case considered in this paper. Such sources should exist for a considerable time: the Ohmic dissipation and some other effects result in the slow diffusion of the magnetic field back to the surface on typical timescales of $10^4$--$10^6$ years \cite{IgoshevEtal2021}. These neutron stars can experience precession. As an example, it follows from Eq.~\eqref{OmegaProp} that the magnetic field $B\sim10^{15}$--$10^{16}$~G, the period of rotation $P\sim0.1$--$1$~s, and the canonical moment of inertia of the neutron star $J=10^{45}\text{ g}\,\text{cm}^2$ imply the period of proper rotation ranging from days to years. Such periodicities might be observed.

\section{Conclusion}

We have constructed a nonradiating electromagnetic source representing a rotating perfectly conducting sphere with a hidden magnetic field. This source is obtained through complete destructive interference of external electromagnetic fields of a uniformly magnetized sphere and a sphere with a nonuniform magnetization that nevertheless provides the same boundary conditions at its surface. The resulting sphere contains a combination of nonzero volume and surface charges and currents the radiation of which is fully self-compensated outside the sphere. These charges and currents are induced through direct mechanical rotation of the sphere and do not require any external electromagnetic antenna-like excitation and using wires, capacitors, or coils. The sphere is not distinguished from the usual nonmagnetized sphere by external radiation or energy dissipation, which is absent, but it has another rotational dynamics. When the axis of rotation is not mechanically fixed, the sphere, though having an isotropic tensor of inertia, executes regular precession without any additional external torque, which at first glance seems counterintuitive as the uniform rotation is expected in this case. The physical reason of the precession is an interesting effect of self-action: rotation of the sphere induces a torque that itself influences the rotation. The torque is due to Lorentz forces appearing from the interaction of internal charges and currents with electromagnetic fields, so the self-action is a sign of generation of electromagnetic fields inside the nonradiating source. This work establishes a link between the electromagnetic and mechanical behavior of nonradiating systems with internal confinement of electromagnetic energy.

\bibliographystyle{elsarticle-num}
\bibliography{paper}

\end{document}